%
%
\def\gap {\vskip 6pt}
\def\indenton {\parindent=20pt}
\def\indentoff {\parindent=0pt}
\def\ref#1#2#3#4#5#6{[#1] #2, {\sl #3}, {\bf #4}, #5, (#6).\vskip 6pt 
\par}
\magnification=\magstep1
\vsize=24 true cm
\goodbreak
\indenton
\vskip 1cm
\centerline {\bf SCREENING OF CHARGED SINGULARITIES OF RANDOM FIELDS}
\vskip 1cm
\centerline {Michael Wilkinson}
\vskip 1cm
\centerline {Department of Applied Mathematics,}
\centerline {Open University,}
\centerline {Walton Hall,}
\centerline {Milton Keynes, MK7 6AA,}
\centerline {England, U.K.}
\vskip 1cm
\vskip 2cm
\centerline{\bf Abstract}
\gap
Many types of point singularity have a topological index, 
or \lq charge', associated with them. For example the phase 
of a complex field depending on two variables can either increase
or decrease on making a clockwise circuit around a simple zero,
enabling the zeros to be assigned charges of $\pm 1$. In random
fields we can define a correlation function for the
charge-weighted density of singularities. In many 
types of random fields, this correlation function
satisfies an identity which shows that the singularities
\lq screen' each other perfectly: a positive singularity is 
surrounded by an excess of concentration of negatives which
exactly cancel its charge, and vice-versa.
This paper gives a simple and widely applicable derivation 
of this result. A counterexample where screening is incomplete
is also exhibited.
\vfill
\eject
%
%
%
%
\noindent{\bf 1. Introduction}
\gap
This paper provides a simple and general explanation
for a feature which has been noted in point singularities
of several types of randomly defined functions.
The phenomenon is most easily described in terms of a specific
example. Consider the set of zeros of a complex valued random 
function $\phi$, depending on two real variables ${\bf x}=(x,y)$. 
It is assumed that the statistical properties of $\phi({\bf x})$ 
are translationally invariant. 
The phase of $\phi$ may either increase or decrease by
$2\pi$ on traversing a clockwise circuit about a simple zero.
Accordingly, the zeros may be described as carrying
either positive or negative charges. It has been noted
that positive zeros tend to be surrounded by negative ones,
and vice-versa: by analogy with models of ionic fluids and
plasmas this has been described as a \lq screening' effect.
The effect is expressed quantitatively in terms of a 
correlation function $C$ describing the charge-weighted
density of zeros (a precise definition will be given in 
section 2). Perfect screening is characterised by
the relation
$$\int d{\bf X}\ C({\bf X})=0
\eqno(1.1)$$
where the integral extends over the entire two-dimensional
plane.

A screening relation of this type was first noted by 
Halperin [1] as a consequence of an analytical evaluation
of such a correlation function in the case of zeros of
complex functions. The screening relation was also 
discussed by Liu and Mazenko [2] in a related context. 
Later, the effect was surmised
to exist in degeneracies of a random matrix model used to
investigate the quantised Hall effect [3].
Numerical demonstrations of perfect screening for
both the zeros and extrema of Gaussian random fields
in two dimensions have been published [4]. 
More recently, analytical evaluation of correlation
functions for several types of point singularity
has provided further evidence that (1.1) is valid
in many cases [5-7]. No general explanation
appears to have been published before now. 

The explanation given here is very simple and widely applicable, and 
leads one to expect that the perfect screening described by (1.1) 
may be present wherever there is a charge-neutral gas of point 
singularities. It depends upon an assumption about another 
correlation function, which is verified in a very 
general context for random fields with a Gaussian distribution, and which
is expected to be much easier to establish than (1.1) in other cases.
One exception to (1.1), relating to components of eigenvectors 
of random matrices, is described and explained.
This counterexample is of some physical interest 
because it arises in the topological description
of the integer quantised Hall effect, discussed in
references [8-10].

This paper will use $\langle A\rangle$ to denote the ensemble 
average of a quantity $A$. Because of the assumption that the statistics
are translationally invariant, the correlation function 
between $\phi({\bf x})$ and $\phi({\bf x}')$ is a function
of ${\bf X}={\bf x}-{\bf x}'$ only. The correlation function may
therefore be written 
$\langle \phi({\bf x})\phi({\bf x}')\rangle=c({\bf x}-{\bf x}')$.
It is distinct from that occuring in (1.1), which describes the singularities
of the field.
\gap\gap\gap
%
%
%
\noindent{\bf 2. A derivation of the screening relation}
\gap
Equation (1.1) will be derived for the case mentioned in the 
introduction, namely zeros of a complex function
of two real variables, $\phi(x,y)$, but the approach is
easily generalised. The total charge
enclosed by a circuit ${\cal C}$ is
$$Q={1\over{2\pi}}\int_{\cal C}{\rm d}s\ 
{\rm Im}\biggl[{1\over {\phi}}{{\rm d} \phi\over{{\rm d}s}}\biggr]
=\int_{\cal C}{\rm d}s\ F(s)
\eqno(2.1)$$
where ${\rm d}s$ is an element of distance around the boundary. Here 
$F(s)$ is shorthand for the integrand: for other types
of singularity where their number can be determined from a line integral 
$F(s)$ would be replaced by a different function. 
It will be assumed that in the more general case 
$F(s)$ is a \lq local' function of
$\phi$, depending only on the function and its derivatives at $s$.
The integral can only be different from zero if the region 
${\cal A}$ enclosed by ${\cal C}$
contains at least one zero of $\phi$. The zeros 
are located at positions ${\bf x}_i$. 
The total charge is
$$Q=\sum_{\{i\vert {\bf x}_i\in {\cal A}\}}q_i
\eqno(2.2)$$
where $q_i$ are the charges of the individual zeros, which 
take the values $\pm 1$.
A charge-weighted density of zeros, $\rho({\bf x})$, is defined:
$$\rho({\bf x})=\sum_i q_i \delta({\bf x}-{\bf x}_i)
\ .
\eqno(2.3)$$
The total charge $Q$ enclosed by ${\cal C}$ can also be expressed 
in terms of this density:
$$Q=\int_{\cal A}{\rm d}A\ \rho({\bf x})
\ .
\eqno(2.4)$$
Because it was assumed 
that the random process generating $\phi({\bf x})$
is translationally invariant, correlation functions 
depend only upon the difference between coordinates. 
We consider random processes which are symmetric between
positive and negative charges, so that 
$\langle \rho ({\bf x})\rangle=0$. We define the mean density 
of zeros (without charge weighting) to be $\rho_0$.
The correlation
function that we consider, already referred to in (1.1), is defined by
$$C({\bf X})=\langle \rho({\bf x}+{\bf X})\rho({\bf x})\rangle
\ .
\eqno(2.5)$$
This correlation function is the sum of a singular part,
$\rho_0\delta({\bf X})$, and a regular part, which 
according to (1.1) is expected to cancel the weight of the
delta function when integrated over all ${\bf X}$.

Now consider the correlation function of the quantity
$F(s)$ introduced in (2.1). The correlation between
$F(s)$ and $F(s')$ depends upon both the distance between
the points labelled by $s$ and $s'$, and on the angle between
the tangent vectors to the curve ${\cal C}$ 
at these points. Within broad limits 
the choice of the curve ${\cal C}$ is immaterial to the argument,
so that this may be taken to be a circle of radius $R$.
In this case:
$$\langle F(s)F(s')\rangle=D(s-s',R)=D_0(s-s')+O(1/R)
\eqno(2.6)$$
where $D_0$ is the limiting form of the correlation function
which applies when the tangent vectors at $s$ and $s'$ 
are parallel.

In order to prove (1.1) it will be assumed that the correlation functions 
$D_0(s-s')$ decays faster than $1/\vert s-s'\vert$
as $\vert s-s'\vert \to \infty$. If $\phi$ is a Gaussian random
function, it is possible to show that this condition is satisfied if 
the magnitude of the correlation 
$c({\bf X})=\langle \phi({\bf X}+{\bf x})\phi({\bf x})\rangle$
decreases sufficiently rapidly as $\vert {\bf X}\vert \to \infty$.
The argument is described in an Appendix; it is applicable to 
other types of point singularities as well as zeros of complex 
functions. 

Using (2.1) and (2.6), $\langle Q^2\rangle$ is expressed 
in terms of the correlation function of $F$:
$$\langle Q^2\rangle=\int_{\cal C}{\rm d}s\int_{\cal C}{\rm d}s'\ 
D(s-s',R)
\ .
\eqno(2.7)$$
Assuming that the correlation function $D_0$ decays
faster than $1/\vert s-s'\vert$ as $\vert s-s'\vert \to \infty$,
in the limit where $R$ is large, the integral (2.7) is dominated
by the region where $s-s'$ is small:
$$\langle Q^2\rangle=2\pi R\int_{-\infty}^\infty {\rm d}s\ D_0(s)
+O(1)
\ .
\eqno(2.8)$$
To gain information about the correlation function
$C({\bf X})$, consider the mean-squared charge: from (2.4)
we have
$$\langle Q^2\rangle=\int_{\cal A}{\rm d}{\bf x}
\int_{\cal A}{\rm d}{\bf x}'\ C({\bf x}-{\bf x}')
\ .
\eqno(2.9)$$
Again, consider the case where ${\cal A}$ is the interior
of a circle radius $R$. If the integral on the left hand side
of (1.1) exists, then the integral in (2.9) is dominated by 
contributions from where ${\bf x}$ is close to ${\bf x}'$, so that
$$\langle Q^2\rangle =\pi R^2\int {\rm d}{\bf X}\ C({\bf X})+O(R)
\ .
\eqno(2.10)$$
Equations (2.8) and (2.10) are only consistent if the term proportional
to $R^2$ in (2.8) vanishes. This implies that (1.1) is satisfied.
This argument has established that (1.1) is correct for the case
of zeros of a complex function.

If $\phi $ is not Gaussian, there does not appear to be any 
general argument which indicates that $D_0(s-s')$ decays 
sufficiently rapidly. The argument given in the appendix can be
adapted to various types of non-Gaussian field, but different 
examples must be treated on a case-by-case basis. 
However, it is expected to be much 
easier to establish that the integral (2.8) converges 
than to establish (1.1), which is an exact identity 
satisfied by a highly singular function of 
the underlying field $\phi$.
\gap\gap\gap
\noindent{\bf 3. Generalisations}
\gap
The argument presented in section 2 may be extended to many
different random fields in different dimensions: there is no 
requirement that it should be possible to determine the correlation 
function $C({\bf X})$ exactly, and the argument can be extended
to non-Gaussian fields.

The argument for perfect screening might be applicable to any 
point singularities which carry indices which we term \lq charges',
when the gas of charges is, on average, neutral. The critical
requirement is that the charge within a closed region can be
determined from a surface integral involving a field $F$ which was
derived in some way from the original random field $\phi$. It was
assumed that this secondary field has a correlation function
which decays sufficiently rapidly at infinity, and which
depends only upon the separation of the pair of coordinates.

Most charged point singularities can be detected by using a
surface integral. As a second example, consider stationary 
points of a real function of two real variables. These may
be characterised by the Poincar\' e index, which is defined
by taking a clockwise circuit around the singularity.
The index is $+1$ if the angle of the gradient vector rotates
by $+2\pi$ (i.e., in the same direction as the circuit), and
$-1$ if the gradient vector rotates in the opposite direction
to the circuit. Thus maxima and minima have index $+1$, and saddles
have index $-1$. Perfect screening has been also been 
demonstrated in this case by calculating the correlation 
function exactly [1,2], but it is instructive to see how
the argument of section 2 is adapted in this simple case.

If ${\bf x}(s)$ is the closed curve ${\cal C}$, and 
${\bf v}(s)=\nabla \phi({\bf x}(s))$ is the gradient
vector at $s$, then the total Poincar\' e index for the
stationary points within ${\cal C}$ is
$$Q={1\over{2\pi}}\int_{\cal C}{\rm d}s\ 
{  
{\bf v} \wedge {{\rm d}{\bf v}\over{{\rm d}s}}
\over{\vert {\bf v}\vert^2
}}
=\int_{\cal C}{\rm d}s\ F(s)
\eqno(3.1)$$
where the second equality defines $F(s)$ in this example.
All of the arguments that were applied to the function
$F(s)$ defined in (2.1) are equally valid for that
defined in (3.1). It follows that the correlation function
of the density of extrema weighted by their 
Poincar\' e index also exhibits perfect
screening under quite general conditions.

The argument also extends directly to cases in higher
dimensions, for example it can be used to explain the 
example of perfect screening which was discussed in reference
[3]. The reasoning for this case will be summarised briefly: the
reader should refer to the earler papers for definitions
of the quantities. 
Reference [3] considered the density of degeneracies
between pairs of levels in random Hermitean matrices,
which were a periodic function of three parameters, $x_1$, 
$x_2$ and $x_3$. There is a topological charge, the Chern index,
associated with the energy levels of a two-parameter family of Hermitean
matrices which are periodic in two parameters, $x_1$ and $x_2$.
The Chern index is an integer which may represent a quantised
Hall conductance associated with each energy level [8,9].
Varying the third parameter $x_3$ allows pairs of levels to become
degenerate. When two levels become degenerate, the Chern 
index on one level increases by one, while that of the other level
decreases by one. This enables the degeneracies to be assigned
a charge of $\pm 1$, depending upon the sign of the change of the
Chern index of the upper level resulting from increasing $x_3$ [10].
Reference [10] also shows that the total charge of the degeneracies 
of a given level (with label $n$, say) within a closed region 
of the three parameter space is equal to the integral of the Berry 
phase two-form $V_n$ [11] over the surface of the region. 

The random matrix model discussed in reference [3] has statistical
properties which are translationally invariant 
in ${\bf x}=(x_1,x_2,x_3)$ space, and this 
symmtetry means that degeneracies are equally likely to have 
either sign. This is analogous to the situation described in section
2: we have a homogeneous distribution of \lq particles' 
(degeneracies) which are equally likely to have positive 
and negative charges, and the total charge $Q$ within a three
dimensional region is obtained by integrating a function,
the Berry phase two-form, over its surface. Considereing a spherical
volume of radius $R$, and writing $\langle Q^2\rangle$
in terms of the correlation function of charge density,
gives an expression analogous to (2.8), with the leading term
being $4\pi R^3/3$ multiplied by the integral of the charge
density correlation function. Expressing $\langle Q^2\rangle$
in terms of the surface integral leads to an expression 
analogous to (2.10) in
which the leading term is $4\pi R^2$ multiplied by an integral
of the correlation function of the two-form, 
$C({\bf x}-{\bf x}')=\langle V_n({\bf x})V_n({\bf x}')\rangle$.
Provided this correlation function vanishes faster than
$1/\vert {\bf x}-{\bf x}'\vert^2$, that integral converges, and
the correlation function of the charge density satisfies
(1.1) (with the integral now being evaluated over a three-dimensional
space). This argument explains the screening relation noted in 
reference [3].
\gap\gap\gap
\noindent{\bf 4. A counterexample: zeros of eigenvector components}
\gap
The discussion in the previous section indicates that the 
argument explaining perfect screening is very general, and
it might be suspected that perfect screening is universal
in charge-neutral gases of singularities. However this section will
describe a counterexample, which arises from the same random matrix model
as was considered in reference [3]. The mathematical structure 
of the model will be explained, but the reader should refer
to [3] for a discussion of the physical motivation of this
model.

Consider a complex Hermitean $N\times N$ matrix $\tilde H$, with 
elements $H_{nm}(x_1,x_2)$ which are a function of two real 
parameters, $x_1$ and $x_2$. The matrix elements $H_{nm}$
can be Gaussian random functions, depending smoothly
on ${\bf x}=(x_1,x_2)$, with a correlation
function between $H_{nm}({\bf x})$ and $H_{n'm'}({\bf x}')$ 
which depends only upon ${\bf x}-{\bf x}'$ and which decays
faster than $1/\vert {\bf x}-{\bf x}'\vert$ as
$\vert {\bf x}-{\bf x}'\vert\to \infty$. In this case each
component of any eigenvector of $\tilde H$ is a complex random
function of $x_1$ and $x_2$. We consider one eigenvector 
component which will be termed $\phi(x_1,x_2)$. 
The function $\phi(x_1,x_2)$ is regular at $(x_1,x_2)$ unless 
the corresponding eigenvalue is degenerate
at $(x_1,x_2)$. Three parameters
must be varied to cause degeneracies of eigenvalues of a 
complex Hermitean matrix, so that we expect $\phi(x_1,x_2)$ to
be regular everywhere in the $(x_1,x_2)$ plane. 

Let us consider the zeros of $\phi$.
There is nothing to favour one 
index of these zeros over the other, so that the distribution
of zeros across the $(x_1,x_2)$ plane is expected to be charge 
neutral. We can ask whether the 
screening relation applies to the zeros of the eigenvector 
component $\phi(x_1,x_2)$. It will demonstrated that in the case
of $2\times 2$ matrices screeening the screening is not perfect.
It will be argued that
screening is also not perfect when $N>2$.

The indices of zeros of a component of an eigenvector of a two-parameter 
familiy of hermitean matrices play a role in the 
topological characterisation of the integer quantised 
Hall effect [9]. In the case of independent electrons moving
through a perfect two-dimensional crystal, the electron 
states form bands labelled by two Bloch wavevectors,
$x_1$ and $x_2$. The Hamiltonian is a periodic function
of $x_1$ and $x_2$, with a unit cell which is termed the
Brillouin zone. The Hall conductance of a
band is $Qe^2/h$, where $Q$ is an integer. One way to
calculate $Q$ involves looking at any component $\phi$ of the 
eigenvector defining the wavefunction of the band. All of the zeros of
$\phi $ within a Brilloiun zone are located, and 
thier indices $q_i=\pm 1$ are 
determined. The Hall conductance integer $Q$ is then the sum
of the indices: $Q=\sum_i q_i$. 

In cases where the electrons move in a simple periodic
potential the integers $Q$ can be large (although these
situations would be very hard to probe experimentally).
If the system is disordered, or the unit cell is large,
it is reasonable to propose using a random matrix model
for the statistical properties of the Chern numbers. An 
appropriate random matrix model is described in [3].
The mean Chern number must (by symmetry) be equal to
zero: $\langle Q\rangle=0$. The simplest statistical 
characterisation of the Chern numbers is through their
variance $\langle Q^2\rangle$.

The random matrix model in [3] was investigated by a
combination of analytical and numerical approaches.
Numerical experiments reported there support an expression of the 
form 
$$\langle Q^2\rangle=\alpha {\cal A}\sigma^2 \rho^2
\eqno(4.1)$$
where 
$$\sigma^2={\rm det}\biggl[\biggl\langle{\partial E_n\over {\partial x_i}}
{\partial E_n\over{\partial x_j}}\biggr\rangle\biggr]
\eqno(4.2)$$
is a measure of the sensitivity of energy levels to 
perurbation, $\rho $ is the density of states,
${\cal A}$ is the area of the Brillouin zone, and 
$\alpha $ is a constant, determined numerically to be
approximately $0.2$ in the limit where the dimension of 
the matrix is large.
Because equation (4.1) is proportional to ${\cal A}$, perfect
screening does not apply in the case of 
zeros of eigenvector components. We have already seen that
screening is predicted by an argument with a very broad range
of applicability. It is desirable to understand why screening 
fails at two levels. Firstly, which of the criteria stated
in the derivation are not met? And secondly, can it
be seen by an explicit calculation that screening is 
absent?

First, let us consider the reason why the demonstration
presented in section 2 cannot be applied. The argument
uses a function $F(s)$ which yields the phase 
change upon integration around the boundary, and the 
function which was defined there, 
$F(s)=(1/\phi){\rm Im}({\rm d}\phi/{\rm d}s)$,
is also appropriate in this problem. It was assumed
that the function $F(s)$ has a correlation function
which is statistically stationary (such that on the 
circular boundary $\langle F(s)F(s')\rangle$ is a 
function of $s-s'$ only, and that this correlation 
function decays faster than $1/\vert s-s'\vert$.
In the case where $\phi $ is a Gauss random function
with a specified correlation function which is 
statistically stationary and rapidly decreasing,
these assumptions are certainly valid. If $\phi $
is a component of an eigenvector of a random matrix,
these assumptions about $F(s)$ may be challenged.
In order to apply (2.1) we must asume that $\phi(x_1,x_2)$ is a 
smooth function of $x_1$ and $x_2$. The eigenvectors
of a matrix may be multiplied by any complex number
of modulus unity, $\exp[{\rm i}\theta(x_1,x_2)]$ and 
remain eigenvectors. When constructing $\phi(x_1,x_2)$,
from the random matrix $\tilde H(x_1,x_2)$, the 
eigenvectors can be produced by some fixed algorithm
(for matrices of dimension $N>4$, this is necessarily
numerical). This results in a \lq raw' form for $\phi(x_1,x_2)$
which is both statistically stationary and locally 
correlated. However, this function will have discontinuities
in phase which must be removed by multiplying by a 
factor $\exp[{\rm i}\theta (x_1,x_2)]$ (the function 
$\theta (x_1,x_2)$ is not unique). The function
$\theta (x_1,x_2)$ cannot be constructed by any 
locally defined algorithm. It is therefore possible
that the resulting regularised $\phi(x_1,x_2)$ may 
be non-stationary, or non-locally correlated, or
both.

This general argument is a little unsatisfying, because
it does not show that the perfect screening effect must fail.
However the source of the failure can be seen 
clearly in the case of a $2\times 2$
hermitean random matrix, with elements
$$\tilde H({\bf x})=\pmatrix{
f_1({\bf x})&f_2({\bf x})+{\rm i}f_3({\bf x})\cr
f_2({\bf x})-{\rm i}f_3({\bf x})&-f_1({\bf x})\cr
}
\eqno(4.3)$$
where ${\bf x}=(x_1,x_2)$, and we assume that the real-valued 
random functions $f_i({\bf x})$ satisfy
$$\langle f_i({\bf x})\rangle=0
\eqno(4.4)$$
$$\langle f_i({\bf x}+{\bf x}_0)f_j({\bf x}_0)\rangle=
\delta_{ij}C_i(\vert{\bf x}\vert)
\ .
\eqno(4.5)$$
The eigenvalues $E_\pm$ and corresponding normalised 
eigenvectors ${\bf u}_\pm$ are
$$E_\pm=\pm \sqrt{f_1^2+f_2^2+f_3^2}\eqno(4.6)$$
$${\bf u}_\pm=
{\exp[{\rm i}\theta_\pm]\over{\sqrt{(E_\pm-f_1)^2+f_2^2+f_3^2}}}
\pmatrix{
f_2+{\rm i}f_3\cr
E_\pm-f_1\cr
}
\eqno(4.7)$$
where all the variables are real functions of ${\bf x}$, and
where $\theta_\pm ({\bf x})$ is chosen so that the first component
of ${\bf u}_\pm$ is a regular function of ${\bf x}$.
Let us consider the first component of the eigenvector corresponding
to the $E_+$ branch, calling this function $\phi({\bf x})$. Writing
$\psi({\bf x})=f_2({\bf x})+{\rm i}f_3({\bf x})$, we have
$$\phi({\bf x})=
{\exp[{\rm i}\theta]\psi
\over{\sqrt{[\sqrt{f_1^2+\vert \psi\vert^2}-f_1]^2+\vert \psi \vert^2}}}
\ .
\eqno(4.8)$$
Consider the behaviour of $\phi$ at a zero of $\psi$. At each
zero of $\psi$, we have $E_+=\vert f_1\vert$. We must consider
two cases, depending upon whether $f_1$ is positive or negative.
If $f_1$ is negative at the zero of $\psi$, then in the neighbourhood of
this zero we have
$$\phi\sim {\exp[{\rm i}\theta]\psi \over{2f_1}}
\eqno(4.9)$$
so that $\phi$ has a zero with the same index as $\psi $, $\theta $
having no singularity at these points. In the case where $f_1$ is
positive at the zero of $\psi$, in the neighbourhood of this
point we have
$$\phi\sim {\exp[{\rm i}\theta]\psi \over{\vert \psi \vert}}
\ .
\eqno(4.10)$$
At these zeros of $\psi$, $\phi$ does not have a zero, and 
$\theta$ must have a singularity in which it 
increments by $\pm 2\pi$ on making a circuit around the 
zero of $\psi$, in order to cancel the phase
singularity of $\psi/\vert \psi\vert$. 

We can now give a clear picture of why screening is not perfect
in the eigenvector component $\phi$, for the special case of this
$2\times 2$ random matrix model. We have seen that the zeros of
$\phi $ have the same indices as those of $\psi $, which do
exhibit perfect screening. However, only a randomly chosen half of 
the zeros of $\psi$ (selected by the criterion that $f_1$ is
negative at the zero) are represented as zeros of $\phi$.
It is therefore not expected that the delicate balance implied
by perfect screening will be present in the zeros of $\phi$.
In the limit where the correlation length of $f_1({\bf x})$
is made short compared to that of $f_2$ and $f_3$, the deleted
zeros are selected completely randomly.

This conclusion can be expressed quantitatively as follows.
The sign-weighted density of zeros of the eigenvector component 
$\phi({\bf x})$ is 
$$\rho_\phi({\bf x})=\chi(f_1({\bf x}))\rho_\psi({\bf x})
\eqno(4.11)$$
where $\rho_{\psi}({\bf x})$ is the sign-weighted density
of zeros of $\psi=f_2+{\rm i}f_3$, and the factor $\chi(f_1({\bf x}))$ 
selects those zeros for which $f_1({\bf x})$ is negative: $\chi(x)$
is unity if $x$ is negative, zero otherwise. Using the fact
that $f_1$ is independent of $\psi$, we have
$$C_\phi({\bf x})=
\langle \chi(f_1({\bf x}_0))\chi(f_1({\bf x}_0+{\bf x}))\rangle 
C_\psi({\bf x})$$
$$=C_\chi({\bf x})C_\psi({\bf x})\eqno(4.12)$$
where the second equality defines $C_\chi({\bf x})$. 
The function $C_\psi$ satisfies equation (1.1). The 
function $C_\phi$ need not. For example, if $f_1$ has a
correlation length which is short compared to $f_2$ and $f_3$,
the correlation length of $C_\chi$ will also be short 
compared to that of $C_\psi$. Equation (4.12) shows that
in this limiting case the screening effect would be
absent from the eigenvector component $\phi$.
\gap\gap\gap
\vfill
\eject
%
%
%
%
\indentoff {\bf References}
\gap
\indentoff
\gap
[1] B. I. Halperin, {\sl Statistical Mechanics of Topological Defects}, 
in {\sl Les Houches Lectures XXV - Physics of Defects}, 
eds. R. Balian, M. Kl\' eman, J-P. Poirier, Amsterdam: North Holland,
(1981).
\gap
\ref {2}{F. Liu and G. F. Mazenko}{Phys. Rev.}{B46}{5963-71}{1992}
\gap
\ref {3}{P. N. Walker and M. Wilkinson}{Phys. Rev. Lett.}{74}{4055-8}{1995}
\gap
\ref {4}{I. Freund and M. Wilkinson}{J. Opt. Soc. Am.}{A15}{2892-902}{1998}
\gap
\ref {5} {M. R. Dennis}{J. Phys.}{A36}{6283-300}{2003}
\gap
\ref {6} {G. Foltin}{J. Phys.}{A36}{1729}{2003}
\gap
\ref {7} {G. Foltin}{J. Phys.}{A36}{4561}{2003}
\gap
\ref {8}{D. J. Thouless, M. Kohmoto, M. P. Nightingale and M. den Nijs}
{Phys. Rev. Lett.}{49}{405-8}{1982}
\gap
\ref {9}{D. J. Thouless}{Phys. Rev.}{B27}{6083}{1983}
\gap
\ref {10}{B. Simon}{Phys. Rev. Lett.}{51}{2167}{1983}
\gap
\ref {11}{M. V. Berry}{Proc. Roy. Soc. Lond.}{A392}{45-57}{1984}
\vfill
\eject
\indenton
%
%
%
\noindent{\bf Appendix}
\gap
This appendix discusses the calculation of the correlation function
$\langle F(s)F(s')\rangle$, in the limit where the separation
$\vert s-s'\vert$ is large. It is assumed that $F(s)$ is a function
of the field $\phi(s)$ and its derivative $\phi'(s)$, that is 
$F(s)={\cal F}(\phi(s),\phi'(s))$ for some function 
${\cal F}(a,b)$, but the argument is easily extended 
to cases where $F(s)$ depends on higher derivatives.
The argument can also be adapted to higher dimensions. 
The calculation discussed here does use the assumption 
that the field $\phi$ is Gaussian, but extensions to non-Gaussian fields
are possible.

It will be shown that $\langle F(s)F(s')\rangle$
decays no more slowly (as $\vert s-s'\vert \to \infty$) 
than the most slowly decaying of the correlations
$\langle \phi(s)\phi(s')\rangle$, $\langle \phi'(s)\phi'(s')\rangle$, 
$\langle \phi'(s)\phi(s')\rangle$. The correlation functions
of derivatives are related to the derivatives of the correlation
function, for example if $\langle \phi(s)\phi(s')\rangle=c(s-s')$, 
then $\langle \phi'(s)\phi'(s')\rangle=-c''(s-s')$.

The joint probability density for $N$ Gaussian random variables
$(x_1,x_2,..,x_N)^T={\bf X}$, all of which satisfy $\langle x_i\rangle=0$, is
$$\mu({\bf X})=\bigl[(2\pi)^N {\rm det}\tilde C\bigr]^{-1/2}
\exp(-{\textstyle{1\over 2}}{\bf X}^T  \tilde C^{-1} {\bf X})
\eqno(A.1)$$
where $\tilde C$ is the correlation matrix of the random variables
$x_i$, with elements
$$C_{ij}=\langle x_i x_j \rangle
\ .
\eqno(A.2)$$
Now write ${\bf X}=({\bf x}_1,{\bf x}_2)^T$, ${\bf x}_1=(\phi(s),\phi'(s))^T$,
${\bf x}_2=(\phi(s'),\phi'(s'))^T$, and write the correlation
matrix (A.2) in the form
$$\tilde C=
\pmatrix{\tilde A & \epsilon (s-s')\tilde a(s-s')
\cr \epsilon (s-s')\tilde a(s-s')& \tilde A\cr}
\eqno(A.3)$$
where $\tilde A$ and $\tilde a$ are $2\times 2$ matrices;
$\tilde A$ is independent of $s-s'$, $\tilde a$ has elements
which are bounded as $\vert s-s'\vert \to \infty$, 
and $\epsilon$ is a function of $s-s'$ which decays no more
rapidly than the most slowly decreasing correlation function of
$\phi$ and $\phi'$ as $\vert s-s'\vert \to \infty$.

When $\vert \epsilon \vert$ is sufficiently small, the inverse
of $\tilde C$ is approximated by
$$\tilde C^{-1}=\pmatrix{
\tilde A^{-1}&-\epsilon \tilde A^{-1}\tilde a\tilde A^{-1}\cr
-\epsilon\tilde A^{-1}\tilde a\tilde A^{-1}&\tilde A^{-1}\cr
}
+O(\epsilon^2)
\ .
\eqno(A.4)$$
We also have ${\rm det}(\tilde C)=[{\rm det}(\tilde A)]^2+O(\epsilon^2)$.
Using (A.4) and (A.1), the joint probability density 
for ${\bf x}_1$ and ${\bf x}_2$
is
$$\mu({\bf x}_1,{\bf x}_2)={1\over {(2\pi)^2{\rm det}(\tilde A)}}
\exp\bigl[-{\textstyle{1\over 2}}({\bf x}_1,{\bf x}_2)^T\tilde C^{-1}
({\bf x}_1,{\bf x}_2)\bigr]+O(\epsilon^2)$$
$$=\mu_0({\bf x}_1)\mu_0({\bf x}_2)
\exp\bigl[\epsilon {\bf x}_1^T\tilde A^{-1}\tilde a\tilde A^{-1}{\bf x}_2
\bigr]
+O(\epsilon^2)
\eqno(A.5)$$
where 
$$\mu_0({\bf x})={1\over{2\pi \sqrt{{\rm det}\tilde A}}}
\exp[-{\textstyle{1\over 2}}{\bf x}^T\tilde A^{-1}{\bf x}]
\eqno(A.6)$$
is the marginal probability density of ${\bf x}=(\phi,\phi')$.

The correlation function $\langle F(s)F(s')\rangle=
\langle {\cal F}({\bf x}_1){\cal F}({\bf x}_2)\rangle$ is
$$\langle F(s)F(s')\rangle=\int {\rm d}{\bf x}_1 \int {\rm d}{\bf x}_2\ 
\mu({\bf x}_1,{\bf x}_2) {\cal F}({\bf x}_1) {\cal F}({\bf x}_2)
\ .
\eqno(A.7)$$
The expectation $\langle F(s)\rangle$ is given by an analogous
expression, in which ${\cal F}({\bf x})$ is integrated over ${\bf x}$ 
with weight $\mu_0({\bf x})$. 
Expanding the exponential in (A.5), and using the fact
that $\langle F(s)\rangle=0$ gives
$$\langle F(s)F(s')\rangle=
\epsilon \int {\rm d}{\bf x}_1 \int {\rm d}{\bf x}_2\ 
\mu_0({\bf x}_1)\mu_0({\bf x}_2) {\cal F}({\bf x}_1){\cal F}({\bf x}_2)
{\bf x}_1^T \tilde A^{-1}\tilde a
\tilde A^{-1}{\bf x}_2+O(\epsilon^2)$$
$$=\epsilon {\bf g}^T\tilde A^{-1}\tilde a \tilde A^{-1}{\bf g}
+O(\epsilon^2)
\eqno(A.7)$$
where 
$${\bf g}=\int {\rm d}{\bf x}\ {\bf x} {\cal F}({\bf x})\mu_0({\bf x})
\ .
\eqno(A.8)$$
Equation (A.7) shows that the correlation function $\langle F(s)F(s')\rangle$ 
is $O(\epsilon)$, implying that it is bounded by a 
multiple of the most slowly decaying correlation function
of the fields occuring as arguments of the function ${\cal F}$.
It follows that the correlation function of $F(s)$ will decay
sufficiently rapidly at infinity if the correlation function
of the fields has a sufficiently rapid decay.

This approach extends directly to non-Gaussian fields 
in the commonly encountered case where the joint probability
density factorises in the limit $\vert s-s'\vert \to \infty$, such
that $\mu({\bf x},{\bf x}')=\mu_0({\bf x})\mu_0({\bf x}')[1+O(\epsilon)]$.
\vfill
\eject
\end